\documentclass[%
 reprint,
superscriptaddress,
 amsmath,amssymb,
 aps,
]{revtex4-2}

\usepackage{graphicx}
\usepackage{dcolumn}
\usepackage{bm}
\usepackage{hhline}
\usepackage{ulem}
\newcommand{\de}{\mathrm{d}}

\begin{document}

\title{Impact of physical-chemistry on the film thinning in
surface bubbles}

\author{Marina Pasquet}
\affiliation{%
Laboratoire de Physique des Solides, CNRS, Université Paris-Saclay, 91405 Orsay
}%

\author{François Boulogne}
\affiliation{%
Laboratoire de Physique des Solides, CNRS, Université Paris-Saclay, 91405 Orsay
}%
\author{Julien Saint-Anna}
\affiliation{%
Laboratoire de Physique des Solides, CNRS, Université Paris-Saclay, 91405 Orsay
}%

\author{Frédéric Restagno}
\affiliation{%
Laboratoire de Physique des Solides, CNRS, Université Paris-Saclay, 91405 Orsay
}%
\author{Emmanuelle Rio}
\email{Corresponding author: emmanuelle.rio@universite-paris-saclay.fr}
\affiliation{%
Laboratoire de Physique des Solides, CNRS, Université Paris-Saclay, 91405 Orsay
}%

\date{\today}

\begin{abstract}
In this paper, we investigate the thinning dynamics of evaporating surfactant-stabilised surface bubbles by considering the role of the physical-chemistry of solutions used in the liquid bath. We study the impact of the surfactant concentration below and above the cmc (critical micelle concentration) and the role of ambient humidity. First, in a humidity-saturated atmosphere, we show that if the initial thickness depends on the surfactant concentration and is limited by the surface elasticity, the drainage dynamics are very well described from the capillary and gravity contributions. These dynamics are independent of the surfactant concentration. In a second part, our study reveals that the physical-chemistry impacts the thinning dynamics through evaporation. We include in the model the additional contribution due to evaporation, which shows a good description of the experimental data below the cmc. Above the cmc, although this model is unsatisfactory at short times, the dynamics at long times is correctly rendered and we establish that the increase of the surfactant concentration decreases the impact of evaporation. Finally, the addition of a hygroscopic compound, glycerol, can be also rationalized by our model. We demonstrate that glycerol decreases the bubble thinning rate at ambient humidity, thus increasing their stability. 
\end{abstract}

\maketitle


\section{Introduction}

The study of surface bubbles have raised a large research interest due to the broad range of applications. 
For instance, we can cite industrial applications such as the flavor of fizzy drinks \cite{Liger-Belair16545}, or the understanding of the dispersion of pollutants above swimming pools \cite{poulain2018}.
In addition, understanding the bubble bursting phenomenon is crucial for modelling the climate, as the process is involved in the quantification of the gas exchange between the ocean and the atmosphere \cite{murphy1998influence,boucher2013clouds}.

In presence of surfactants, the bubble lifetime is primarily fixed by the thinning velocity \cite{miguet2020stability}, which in turn is due to two mechanisms, the drainage 
and the liquid evaporation
.
Indeed, the film surrounding surface bubbles has been proven to thin down to a few tens of nanometers, the thickness at which bursting occurs. 
To understand and to be able to predict the lifetime of these bubbles, it is therefore necessary to consider their thinning dynamics~\cite{miguet2021life}.

In absence of evaporation, the drainage is driven by the competing effects of capillary suction \cite{lhuissier2012bursting} and gravity \cite{miguet2020stability}.
This process is limited by the flow through a pinch developing at the bottom of the surface bubbles, in the vicinity of the meniscus \cite{Aradian2001}.
Such a description has been verified experimentally, either in a saturated atmosphere, \textit{i.e.} in absence of evaporation \cite{miguet2020stability} or at a short time, when evaporation is negligible over the drainage \cite{lhuissier2012bursting, poulain2018ageing}.

Nevertheless, evaporation must be taken into account for thicknesses typically lower than one micrometer \cite{Champougny2016}. 
To describe the thinning rate in presence of evaporation, a constant evaporation rate can be added to the mass conservation. 
Such consideration provides a successful description of the data available in the literature \cite{poulain2018ageing,miguet2020stability}.

A noticeable consequence of this description lies on the role of the liquid composition, which is expected to modify the thinning process, and thus the bubble lifetime, only through the liquid viscosity, density, and surface tension.
This observation conflicts to the common intuition and daily observations that the choice of the soapy solution recipe is fundamental to control bubbles stability \cite{frazier2020make}, whereas viscosity, density, and surface tension are slightly modified.

Hence, the aim of the paper is to identify the role of physical-chemistry on surface bubble thinning, and in particular to precise separately the specific effects on drainage and on evaporation.
To make this distinction, we performed experiments in a saturated atmosphere to identify the role of the surfactant concentration on thinning dynamics, without any evaporation phenomena.
Then, additional experiments are presented in an atmosphere controlled in humidity to investigate the role of evaporation.

\section{Experimental setup}

\subsection{Material}

In this study, we use TTAB (Tetradecyl Ammonium Bromide, purchased from Sigma-Aldrich) as well as glycerol (purity $\leq$ 99.5 \% purchased from VWR) diluted in ultrapure water (resistivity = 18.2 M$\Omega \cdot $cm).
The TTAB solutions are prepared for several concentrations between 0.5 and 20 times the critical micelar concentration (cmc = 3.6 mmol$\cdot$L$^{-1}$). 
Since impurities in surfactant solutions have a noticeable effect on surface properties below the cmc, TTAB used 0.5 and 1 times the cmc is recrystallized prior the preparation of the solutions \cite{stubenrauch2005foams}.

To investigate the impact of glycerol on the stability of surface bubbles, solutions with TTAB are made by replacing ultrapure water with a mixture of ultrapure water and glycerol with an initial glycerol concentration $c_{g_0} = $ 20 wt\%. 

The surface tensions $\gamma$ of our solutions are measured with a pendant drop commercial apparatus (Tracker, Teclis) and the results are reported in Table \ref{tab:SurfaceTensions}.
The viscosity $\eta$ and the density $\rho$ of the solutions containing TTAB in water are assumed to be identical to the pure water properties.
The viscosity and density of solution containing glycerol are estimated from data on water-glycerol mixtures available in the literature \cite{cheng2008formula,volk2018density}.

\begin{table}
\centering
\begin{tabular}{|c|c|c|c|c|} 
\hline
system    & \multicolumn{1}{c}{\begin{tabular}[c]{@{}c@{}}$c$\\ \end{tabular}} & \multicolumn{1}{c}{\begin{tabular}[c]{@{}c@{}}$\gamma$\\(mN/m)\end{tabular}} & \multicolumn{1}{c}{\begin{tabular}[c]{@{}c@{}}$\rho$\\(kg/m$^3$)\end{tabular}} & \begin{tabular}[c]{@{}c@{}}$\eta$\\ (mPa$\cdot$s)\end{tabular}  \\
\hhline{|=====|}
TTAB & 0.5 cmc   & 50.2 &998 & 1.00   \\
TTAB & 1 cmc    & 38.2 &998 & 1.00   \\
TTAB & 5 cmc   & 35.9  &998 & 1.00  \\
TTAB & 10 cmc  & 35.7 &998 & 1.00   \\
TTAB & 20 cmc  & 35.2  &998 & 1.00  \\
TTAB/Gly & 0.5 cmc    & 42.3   &1047 & 1.74 \\
TTAB/Gly & 20 cmc  & 23.4  &1047 & 1.74 \\
\hline
\end{tabular}
    \caption{Surface tension $\gamma$ measured for TTAB solutions at different concentrations noted $c$ in the table, at 20~$^{\circ}$C (at $\pm 0.2$ mN/m).
    TTAB used for the solutions at 0.5 and 1 times the cmc have been recrystallized. The denomination "TTAB/Gly" corresponds to solutions containing a mass concentration of glycerol of 20 \%. }
    \label{tab:SurfaceTensions}
\end{table}
\subsection{Bubble generation and thinning measurement}

\begin{figure}[ht]
  \centering
    \includegraphics[width=\linewidth]{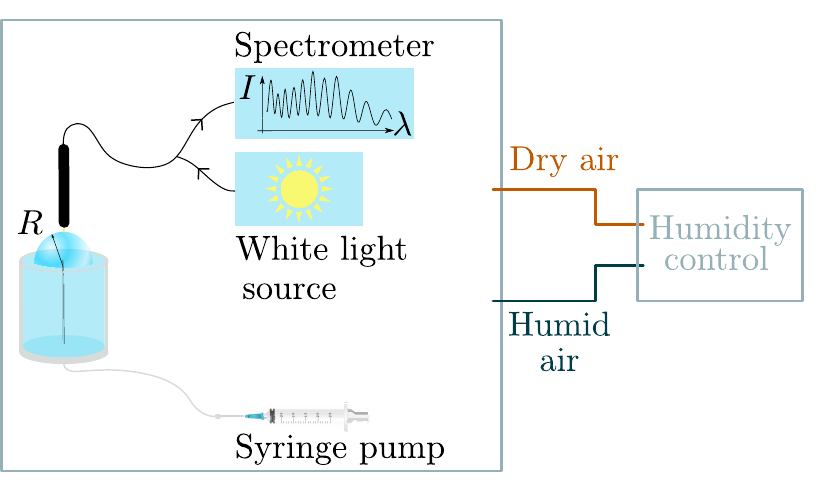}
  \caption{A container with the solution of interest is set in a humidity-controlled chamber. 
  To generate the bubble, air is injected by a syringe pump. 
  The film thickness at the apex of the surface bubble is measured by white light spectrometry.}
  \label{fig:Setup}
\end{figure}

The experimental setup is schematized in Fig. \ref{fig:Setup}.
Bubbles are generated on the surface of a cylindrical Plexiglas tank of 4.0 cm diameter and 4.6 cm depth, filled with a solution up to the rim. 
To inject air in the solution, a PTFE tubing with an external diameter of 0.56 mm (from Cole-Parmer) is inserted at the base of the tank. 
This extremity of the tube is beveled and placed just below the surface of the solution. 
To circumvent the flexibility of the PTFE tubing, the later is inserted in a glass capillary tube.
The opposite extremity of the PTFE tubing is then connected to a flexible tube of 6 mm diameter, and about 80 cm long connected to a 50 mL syringe. 
The syringe is placed on a syringe pump (AL-1000 from WPI), such that the airflow rate $Q$ and the injected volume $V$ are perfectly controlled. 
In this study, we choose to keep the injected volume $V = 1.50$ mL and the flow rate $Q = 15$ mL$\cdot$min$^{-1}$ constant, such that the bubble radius of curvature $R = 1.03 \pm 0.08$ cm is also constant.

The temperature is monitored and ranges between 20 and 23~$^{\circ}$C. 
The whole device is enclosed in a chamber of size $40 \times 40 \times 50$ cm$^{3}$.
The humidity is controlled through a humidity regulator \cite{Boulogne2019}. 
For experiments conducted at 100 \% humidity, a humidifier (Bionaire BU1300W-I) and wet towels are also used to saturate the box with water vapor.

The thickness $h$ of the film at the apex of the bubble cap is measured using a UV-VIS spectrometer (Ocean Optics Nanocalc 2000), whose wavelengths are between 450 and 800~nm, associated with a 200 $\mu$m diameter optical fiber. 
The spectrometer must be positioned vertically, precisely at the apex to allow the spectrometer to be perpendicular to the film. 
The tank containing the solution is thus placed on a $xy$-translation plate. 
The thickness is extracted from the spectrum of the reflected light using the \textit{Oospectro} \cite{Oospectro} library written in Python \cite{miguet2020stability, MiguetPRF2021}.
The temporal evolution of the film thickness is measured two or three times for each experimental condition.
As the measurements are perfectly reproducible, the results presented will correspond systematically to two or three repetitions of the same experiment condition and for better readability, the symbols will be the same. 
The initial time ($t = 0$) corresponds to the end of the bubble generation.

\section{Impact of physical-chemistry on thin film drainage}


 %
\begin{figure}[ht]
  \centering
    \includegraphics[width=\linewidth]{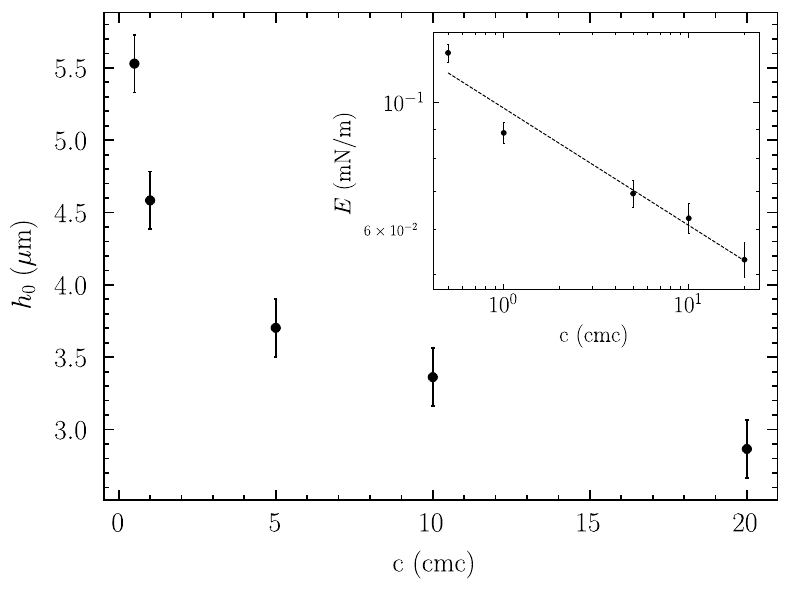}
  \caption{Variation of the initial thickness of surfactant solutions as a function of the TTAB concentration. The error bars correspond to the estimated accuracy of 0.2~$\mu$m on the determination of the initial thickness.
  The inset shows the Gibbs-Marangoni elasticity of the film obtained from $h_0$ using Eq. \ref{eq:InitialThickness}.  
  The dashed line is a power law fit: $E = 0.10\,c^{-0.20}$.
} 
  \label{fig:InitialThickness}
\end{figure}
\subsection{Initial thickness}
In this Section, we consider the experiments performed in a saturated atmosphere, so that we expect the thinning to be due solely to drainage since no evaporation occurs.
Once the bubble is generated at the surface of the surfactant solution, we measure the initial thickness $h_0$ at the apex.
These measurements are presented in figure~\ref{fig:InitialThickness}, which shows a decrease of the initial thickness with the surfactant concentration.

To understand the variation of $h_0$ with the surfactant concentration, a first approach is to adopt the description of the film generation based on a Frankel-type mechanism \cite{Frankel1959}. 
In this model, the thickness is set by a balance between the capillary suction in the meniscus and the viscous dissipation in the thin film, characterized by the capillary number $\text{Ca} = \eta V /\gamma$ with the entrainment velocity $V$.
The Frankel description leads to an initial thickness $h_0\sim \ell_c\,\text{Ca}^{2/3}$, where $\ell_c = \sqrt{\gamma/(\rho g)}$ is the capillary length, with $g$ the gravitational acceleration.

In the present study, all bubbles are generated with the same flow rate, \textit{i.e.} the same entrainment velocity $V$. 
Additionaly, the surfactant concentration affects mainly the surface tension $\gamma$ (table \ref{tab:SurfaceTensions}). 
The difference of surface tension between TTAB at 0.5 and 20 cmc is of 15 mN/m. 
This leads to a difference of 0.4 $\mu$m for $h_0$, much lower than what is measured. 
As a consequence, this model predicts a nearly identical initial thicknesses for all bubbles based on Frankel's mechanism.
This prediction is in contradiction with the observations presented in figure~\ref{fig:InitialThickness}.

Therefore, we conclude that the soap film generation is not driven by a viscous flow, but rather by elasticity \cite{champougny2015surfactant,seiwert2014theoretical}.
In an elasticity-dominated regime, the thickness is limited by the surface elasticity $E$, \textit{i.e.} by the ability of the surface to sustain surface concentration gradients and thus surface tension gradients.

In this regime, the balance between the capillary suction and the Marangoni stress due to surfactant surface concentration inhomogeneities at the interfaces leads to \cite{champougny2015surfactant,seiwert2014theoretical}:
\begin{equation}
    h_0 \ \sim \ \ell_c \frac{E}{\gamma}.
    \label{eq:InitialThickness}
\end{equation}

With this prediction, the initial thickness is expected to depend on the surfactant concentration through the surface elasticity $E$. 
Here, we propose to deduce the surface elasticity from the initial thickness measurements and comment the result.
From our data $h_0(c)$ and Eq.~\ref{eq:InitialThickness}, we plot in the insert of Fig. \ref{fig:InitialThickness} the predicted surface elasticity.
We obtain a decay of the surface elasticity with the surfactant concentration, which we describe with a power law model. 
Such decrease is expected since a larger surfactant concentration allows a faster repopulation of the interfaces and thus a smaller surface elasticity \cite{langevin2014rheology}. 
The measurement of surface elasticity is delicate and rare in the literature. In our study, the order of magnitude of the surface elasticity is 0.1 mN/m, which is lower but in agreement with those measured by Champougny \textit{et al.} \cite{champougny2015surfactant}
(C$_{12}$E$_{6}$), which are between 0.15 and 0.59 mN/m.
This non-ionic surfactant is much less soluble that TTAB.
In such a situation, the surface elasticity is expected to be slightly higher, in agreement with our experiment.

%
\begin{figure}[!ht]
  \centering
    \includegraphics[width=\linewidth]{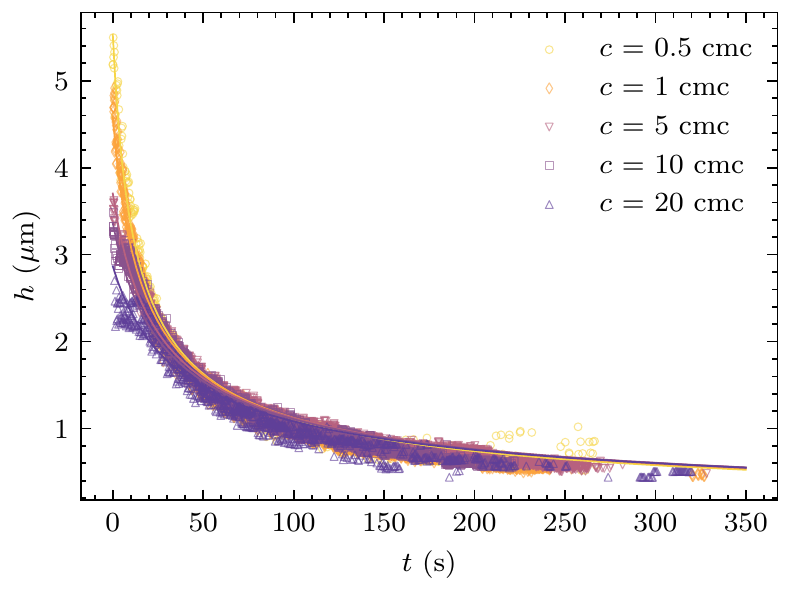}
  \caption{Film drainage in a saturated environment for bubbles stabilized by different concentrations of TTAB ranging in 0.5-20 cmc. 
  The continuous lines correspond to Eq. \ref{eq:Drainage}.  }
  \label{fig:TTAB_RH100}
\end{figure}

\subsection{Thinning dynamics}
Now that the initial thickness is described, we focus our attention on the thinning dynamics.
All the drainage curves obtained for the different TTAB concentrations are presented in Fig. \ref{fig:TTAB_RH100}.
Remarkably, the thinning process appears to be nearly independent on the surfactant concentration, besides the initial thickness described in the previous paragraph.

In the following, we compare these thinning dynamics with the drainage model described in the introduction, in which the flow is limited by the presence of a pinch in the vicinity of the contact line \cite{lhuissier2012bursting, poulain2018ageing, miguet2020stability}.
The thinning rate is therefore the combination of capillary-driven and gravity-driven flows, which reads

\begin{equation}
    \dfrac{\de h}{\de t} \  = \ - \ \dfrac{\gamma}{\eta} \left( \dfrac{h^{5/2}}{R^{5/2}} + \dfrac{h^3}{R \ell_c^2} \right) ,
    \label{eq:Drainage}
\end{equation}

Indeed, the gravity-driven drainage is relevant in our situation because the Bond numbers $\text{Bo}$ comparing the effect of gravity and surface tension ($\text{Bo} = \rho g R^2/\gamma$) corresponding to our bubbles are between 20 and 44.  
To solve this differential equation we used the \textsf{odeint} function from the scipy module \cite{Jones2001} in Python.
The initial thickness of the bubbles $h_0$ is left as an adjustable parameter for the integration of equation \ref{eq:Drainage}, \textit{via} the use of the \textsf{curve\_fit} function of the scipy module \cite{Jones2001}. 
This thickness is plotted as a function of TTAB concentration in Figure \ref{fig:InitialThickness}.
Figure \ref{fig:TTAB_RH100} shows that the experimental data agree with this theoretical prediction (solid lines).

Finally, our first conclusion is that, in absence of evaporation, the concentration in TTAB does not impact the drainage dynamics.
This is confirmed by the results of Poulain \textit{et al} \cite{poulain2018ageing} for bubbles of radius $R = 4.8 \pm 0.1$ mm with another surfactant, SDS (Sodium Dodecyl Sulfate). 
The only impact of the chemical-chemistry is a short-time effect. 
The liquid composition changes the initial thickness through the variation of the surface elasticity.
After a few tens of seconds, which is short compared to the bubble draining time, all the dynamics becomes independant of the TTAB concentration.
We thus expect a negligible impact on the bubble lifetime.

\section{Impact of physical-chemistry on thin film evaporation}

\subsection{Control of evaporation via the atmospheric humidity}

To explore the impact of physical-chemistry on evaporation, we now measure the thinning dynamics at relative humidities $\text{RH}$ equal to 39 and 55 \%.

\begin{figure}
  \centering
    \includegraphics[width=\linewidth]{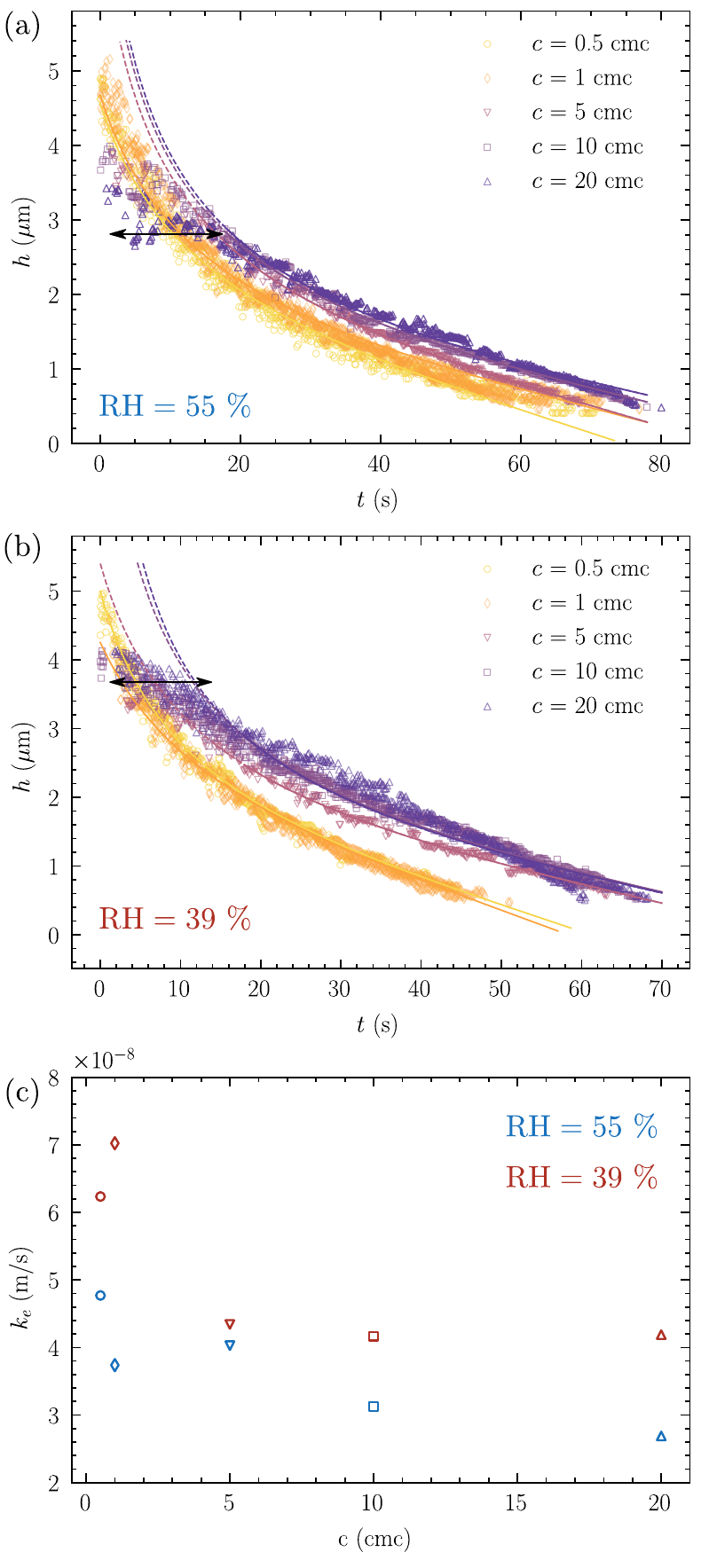}
  \caption{Film thinning at the bubble apex in a environment of controlled humidity of (a) $RH = 55 \pm 4$ \% and (b) $RH = 39 \pm 1$ \%, for bubbles stabilized by different concentrations of TTAB ranging in 0.5-20 cmc. The continuous lines correspond to Eq. \ref{eq:DrainageEvap} using $k_e$ (given in (c)) as a fitting parameter. For c $>$ cmc, the fit was made from $t = 18$~s: the extrapolation to the first instants is indicated by a dotted line.   }
  \label{fig:TTAB_RH39-50}
\end{figure}

As expected, the dynamics is much faster in presence of evaporation so that the bubble lifetime is more than 5 times shorter.
Additionally, in conditions for which evaporation proceeds, the thinning dynamics for different surfactant concentrations do not collapse in a single curve as shown in  Fig. \ref{fig:TTAB_RH39-50}.
This observation contrasts with the results in a saturated atmosphere obtained in the previous Section.
Thus, although the surfactant concentrations studied in this article have a little impact on the drainage dynamics, this concentration affect significantly the thinning of the bubbles when evaporation proceeds.

To describe the thinning dynamics, we add an evaporation rate\cite{poulain2018ageing,miguet2020stability} to the former drainage equation \ref{eq:Drainage}.
For this, we can consider a diffusive evaporation rate \cite{Fuchs1959}, which is proportional to the difference in pressure between the saturation vapour pressure at the interface $p_{\text{sat}}$ and in the surrounding environment $p_{\infty}$. 
We write the contribution of the evaporation on the film thinning as $k_e (1-\text{RH})$, where $k_e$ has the dimension of a velocity.
This coefficient contains the geometry of the system, kept constant in this study, the diffusion coefficient of water in air, and the saturated vapor pressure of the chemical solution.
This leads to the thinning rate
\begin{equation}
    \dfrac{\de h}{\de t} \ = \ - \ \dfrac{\gamma}{\eta} \left( \dfrac{h^{5/2}}{R^{5/2}} + \dfrac{h^3}{R \ell_c^2} \right) \ - \ k_e \  (1-\text{RH}).
    \label{eq:DrainageEvap}
\end{equation}


The model describes very well the data for TTAB concentration smaller or equal to the cmc for both $\text{RH}$ = 39 and 55 \%. 
For the solutions containing TTAB above the cmc, the proposed model does not allow to account for the beginning of the thinning curves. 
We observe that, in presence of evaporation, the thinning dynamic is delayed at the beginning for these concentrations, especially for the highest concentration (20 times the cmc) where a plateau, marked by a black arrow, can be observed.
Thus, a fit for these "high" concentrations is proposed only at times greater than 18 seconds (Fig. \ref{fig:TTAB_RH39-50} (a) and (b)). 
In this case, there is a good agreement between our model and the experimental data. 
It can be noted that for concentrations above cmc, the higher the TTAB concentration, the slower the evaporative thinning dynamics, as indicated by the measured $k_e$ constant values given in the Fig. \ref{fig:TTAB_RH39-50} (c).

Our second result is thus that the variation of the TTAB concentration affects the thinning dynamics of surface bubbles through the evaporation rate.
A simple model, in which a constant evaporation flux is added to the drainage dynamics allows describing the data in the first order.

\subsection{Control of evaporation \textit{via} the addition of an hygroscopic component}

\begin{figure}
  \centering
    \includegraphics[width=\linewidth]{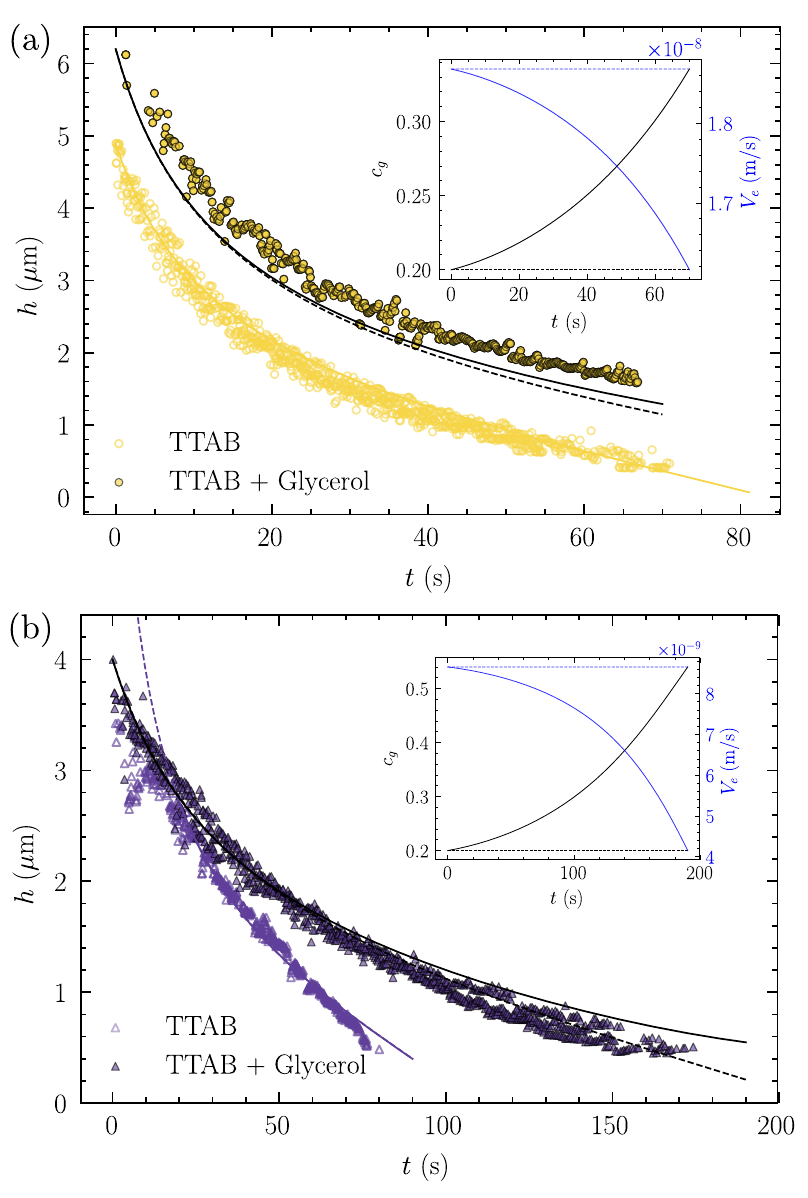}
  \caption{Time evolution of the film thickness $h$ at the bubble apex in a environment of controlled humidity ($55 \pm 5$ \%) for bubbles stabilised with or without an initial glycerol concentration $c_{g_0} = 20$ wt.\%, for fixed TTAB concentrations: (a) 0.5 cmc and (b) 20 cmc. The data in the absence of glycerol and the corresponding colour adjustment correspond to those shown in Fig. \ref{fig:TTAB_RH39-50}. For the data in the presence of glycerol, the black dotted line corresponds to the Eq. \ref{eq:DrainageEvapAds} with a constant concentration $c_g = c_{g_0}$. The black solid line corresponds to this equation taking into account the variation of $c_g$ with time, by combining Eq. \ref{eq:mg}, \ref{eq:mw} and \ref{eq:DrainageEvapAds}.
  The inset shows the variation of $c_g$ and $V_{e}$ for these two cases.}
  \label{fig:ttab_glycerol_RH=50}
\end{figure}

In the following, the surface bubbles studied are generated at RH = 55 $\pm$ 5 \% in water-glycerol mixtures for two concentrations of the TTAB solution (equal to 0.5 and 20 times the cmc). 

The thinning curves obtained for bubbles stabilized by TTAB/glycerol/water solutions, with an initial glycerol concentration $c_{g_0} = $ 20 wt.\%,  are plotted in Fig. \ref{fig:ttab_glycerol_RH=50}.
Comparing these results to those obtained with TTAB/water only, we can see that the addition of glycerol stabilizes the bubbles by decreasing their thinning rate at long times (in particular for the concentration corresponding to 20 times the cmc).

To rationalize this observation, we propose to take into account the addition of glycerol in the thinning dynamics.
Glycerol is a hygroscopic component, known to affect the solution activity\cite{Bouchaudy2018} and thus, its evaporation rate \cite{glycerine1963physical,chen2009hydrogen}.
The saturation vapour pressure $p_{\text{sat}} (c_g) $ at the evaporating interface is modified by the presence of glycerol in the solution and can be written as \cite{glycerine1963physical}: 

\begin{equation}
    p_{\text{sat}} (c_g) \ = \  p_{\text{sat}} (c_g = 0) \dfrac{1 - c_g}{1 + c_g (a-1) }
    \label{eq:psatGly}
\end{equation}
\noindent
with $a = 0.248$ \cite{roux2021bubbles, glycerine1963physical}. 


As in equation \ref{eq:Drainage}, the drainage caused by the capillary suction and gravity remains unchanged.
Only the evaporation term is modified by taking into account the change in saturation vapour pressure at the interface given by equation \ref{eq:psatGly}, which leads to writing the thinning rate such as

\begin{equation} 
\dfrac{\de h}{\de t} \ = \  - \ \dfrac{\gamma}{\eta} \left( \dfrac{h^{5/2}}{R^{5/2}} + \dfrac{h^3}{R \ell_c^2} \right) 
     -  k_e \left( \dfrac{1 - c_g}{1 + c_g (a-1)}-\text{RH} \right) .
    \label{eq:DrainageEvapAds}
\end{equation}
The thinning rate corresponding to evaporation, the second term on the right hand side of this equation, can therefore be positive or negative, which means that in the presence of glycerol, the water can either evaporate (positive term) or condense (negative term). 
This depends on both the glycerol concentration and the relative humidity. 

In this section, we will not consider any adjustable parameter to integrate the equation \ref{eq:DrainageEvapAds}: the initial thickness $h_0$ corresponds to the experimentally measured value and the evaporation constant $k_e$ to the value determined in the previous paragraph, in the absence of glycerol. 
The prediction of Eq. \ref{eq:DrainageEvapAds}, for a constant glycerol concentration $c_g (t) = c_{g_0}$, is shown in black dotted lines in Fig. \ref{fig:ttab_glycerol_RH=50}.

The small discrepancy can be explained in part by the fact that the mass concentration $c_g$ changes along time.
To integrate Eq. \ref{eq:DrainageEvapAds}, the time evolution of the glycerol concentration $c_g(t)$ must be precised from the exchanges with the bath and the atmosphere.
We consider that the bubble has a spherical cap of volume $\Omega(t) = 2/3 \pi (3 R^2 h(t)+3Rh(t)+h(t)^3)$, for which the liquid composition is homogeneous.
The weight of glycerol in the bubble is defined as $m_g(t) = c_g(t) \rho \Omega(t)$.
From the definition of the glycerol concentration $c_g(t) = m_g(t) / (m_g(t) + m_w(t))$, we have $m_w(t) = \left(1-c_g(t) \right) \rho \Omega(t)$.
The time variation of the glycerol weight in the spherical cap is due to drainage of velocity $V_d[h(t)]=\gamma/\eta \left( h^{5/2}/R^{5/2} + h^3/(R \ell_c^2) \right) $, which gives
\begin{equation}\label{eq:mg}
    \frac{{\rm d} m_g}{{\rm d}t} = - \rho_g V_d[h(t)] S \Phi_g(t),
\end{equation}
where $S= 2 \pi R^2$ is the surface area of the hemispherical cap and  $\Phi_g(t) = \left(1+ \rho_g/\rho_w (1/c_g(t)-1)\right)^{-1}$ is the volume concentration of glycerol.
Similarly, for the water content, the variations is due to the drainage velocity $V_d$ but also the evaporation velocity $V_{e}[c_g(t)]= k_e \left( (1 - c_g)/(1 + c_g (a-1))-\text{RH} \right) $, such that
\begin{equation}\label{eq:mw}
    \frac{{\rm d} m_w}{{\rm d}t} = - \rho_w V_d[h(t)] S \Phi_g(t) - \rho_w V_{e}[c_g(t)] S.
\end{equation}

We solve equations \ref{eq:DrainageEvapAds}, \ref{eq:mg} and \ref{eq:mw} for the following initial conditions, using an Euler-type method: $c_{g_0} = 0.2$, $m_{w0} = (1-c_{g_0}) \rho \Omega_0$ and $m_{g0} = c_{g_0} \rho \Omega_0$ with $\Omega_0 = 2/3 \pi (3 R^2h_0+3Rh_0+h_0^3)$ the volume of the liquid in the bubble cap at $t=0$. Note that the surface tension (measured) remains almost constant up to $c_g = 0.85$, so we used the value given in the Tab. \ref{tab:SurfaceTensions}.


The results of this integration are shown in Fig. \ref{fig:ttab_glycerol_RH=50} with black solid lines, describing relatively well the temporal evolution of the measured thickness for the two concentrations. 
The inset shows the evolution of the glycerol concentration and the evaporation rate during the thinning of the bubbles.
As time goes by, the glycerol concentration increases, which leads to a decrease in the evaporation rate and therefore the bubble thinning rate. 
We attribute  the enhanced stability of these bubbles at longer times to the reduction of the thinning rate. 


\section{Conclusion}

In absence of evaporation, physical-chemistry has no impact on the thinning dynamics of surface bubbles.
It affects only the initial thickness, which can be explained by a surface elasticity depending on the surfactant concentration.
In presence of evaporation, the thinning dynamics is affected by the TTAB concentration. 
This effect can be explained at the first order by an evaporation flux depending on the TTAB concentration.
Nevertheless, a more precise description of this dynamic probably requires a better understanding of the role of surfactants on evaporation.
This role can be due to a barrier exerted by the surfactants to evaporation \cite{la2014retardation}, which could explain the variation of the evaporation rate with the surfactant concentration.
Evaporation could also generate additional solutal or thermal surface tension gradients, which could add a stabilizing or destabilizing liquid flow \cite{poulain2018ageing,li2010effect}.
Additionally, the water evaporation could lead to a variation of the local concentration in surfactants, which can also affect the flow and/or evaporation rate. 
This work thus opens new questions, which shall be answered in a future work.
Furthermore, we observed that the addition of glycerol slows down the thinning dynamics. 
This effect can be described by considering both the modification of the saturation vapor pressure in the presence of glycerol and the increasing glycerol concentration in time due to water evaporation.

We have enlighten that the description of surface bubbles thinning dynamics requires a better understanding of the effect of physical-chemistry on evaporation.
This effect can come from different mechanisms (i) the concentration of non volatile components along time \cite{roux2021bubbles} (ii) the existence of a barrier to evaporation due to the presence of surfactants at the interface, which can affect the evaporation flux \cite{la2014retardation} or the apparition of solutal or thermal surface tension gradient \cite{poulain2018ageing,li2010effect}, which can lead to additional liquid velocities. 

\section*{Acknowledgments} Funding from ESA (MAP Soft Matter Dynamics)
and CNES (through the GDR MFA) is acknowledged. 

\bibliography{Biblio} 

\providecommand*{\mcitethebibliography}{\thebibliography}
\csname @ifundefined\endcsname{endmcitethebibliography}
{\let\endmcitethebibliography\endthebibliography}{}
\begin{mcitethebibliography}{29}
\providecommand*{\natexlab}[1]{#1}
\providecommand*{\mciteSetBstSublistMode}[1]{}
\providecommand*{\mciteSetBstMaxWidthForm}[2]{}
\providecommand*{\mciteBstWouldAddEndPuncttrue}
  {\def\EndOfBibitem{\unskip.}}
\providecommand*{\mciteBstWouldAddEndPunctfalse}
  {\let\EndOfBibitem\relax}
\providecommand*{\mciteSetBstMidEndSepPunct}[3]{}
\providecommand*{\mciteSetBstSublistLabelBeginEnd}[3]{}
\providecommand*{\EndOfBibitem}{}
\mciteSetBstSublistMode{f}
\mciteSetBstMaxWidthForm{subitem}
{(\emph{\alph{mcitesubitemcount}})}
\mciteSetBstSublistLabelBeginEnd{\mcitemaxwidthsubitemform\space}
{\relax}{\relax}

\bibitem[Liger-Belair \emph{et~al.}(2009)Liger-Belair, Cilindre, Gougeon,
  Lucio, Gebef{\"u}gi, Jeandet, and Schmitt-Kopplin]{Liger-Belair16545}
G.~Liger-Belair, C.~Cilindre, R.~D. Gougeon, M.~Lucio, I.~Gebef{\"u}gi,
  P.~Jeandet and P.~Schmitt-Kopplin, \emph{Proceedings of the National Academy
  of Sciences}, 2009, \textbf{106}, 16545--16549\relax
\mciteBstWouldAddEndPuncttrue
\mciteSetBstMidEndSepPunct{\mcitedefaultmidpunct}
{\mcitedefaultendpunct}{\mcitedefaultseppunct}\relax
\EndOfBibitem
\bibitem[Poulain and Bourouiba(2018)]{poulain2018}
S.~Poulain and L.~Bourouiba, \emph{Physical Review Letters}, 2018,
  \textbf{121}, 204502\relax
\mciteBstWouldAddEndPuncttrue
\mciteSetBstMidEndSepPunct{\mcitedefaultmidpunct}
{\mcitedefaultendpunct}{\mcitedefaultseppunct}\relax
\EndOfBibitem
\bibitem[Murphy \emph{et~al.}(1998)Murphy, Anderson, Quinn, McInnes, Brechtel,
  Kreidenweis, Middlebrook, P{\'o}sfai, Thomson, and
  Buseck]{murphy1998influence}
D.~Murphy, J.~Anderson, P.~Quinn, L.~McInnes, F.~Brechtel, S.~Kreidenweis,
  A.~Middlebrook, M.~P{\'o}sfai, D.~Thomson and P.~Buseck, \emph{Nature}, 1998,
  \textbf{392}, 62--65\relax
\mciteBstWouldAddEndPuncttrue
\mciteSetBstMidEndSepPunct{\mcitedefaultmidpunct}
{\mcitedefaultendpunct}{\mcitedefaultseppunct}\relax
\EndOfBibitem
\bibitem[Boucher \emph{et~al.}(2013)Boucher, Randall, Artaxo, Bretherton,
  Feingold, Forster, Kerminen, Kondo, Liao,
  Lohmann,\emph{et~al.}]{boucher2013clouds}
O.~Boucher, D.~Randall, P.~Artaxo, C.~Bretherton, G.~Feingold, P.~Forster,
  V.-M. Kerminen, Y.~Kondo, H.~Liao, U.~Lohmann \emph{et~al.}, in \emph{Climate
  change 2013: the physical science basis. Contribution of Working Group I to
  the Fifth Assessment Report of the Intergovernmental Panel on Climate
  Change}, Cambridge University Press, 2013, pp. 571--657\relax
\mciteBstWouldAddEndPuncttrue
\mciteSetBstMidEndSepPunct{\mcitedefaultmidpunct}
{\mcitedefaultendpunct}{\mcitedefaultseppunct}\relax
\EndOfBibitem
\bibitem[Miguet \emph{et~al.}(2020)Miguet, Pasquet, Rouyer, Fang, and
  Rio]{miguet2020stability}
J.~Miguet, M.~Pasquet, F.~Rouyer, Y.~Fang and E.~Rio, \emph{Soft Matter}, 2020,
  \textbf{16}, 1082--1090\relax
\mciteBstWouldAddEndPuncttrue
\mciteSetBstMidEndSepPunct{\mcitedefaultmidpunct}
{\mcitedefaultendpunct}{\mcitedefaultseppunct}\relax
\EndOfBibitem
\bibitem[Miguet \emph{et~al.}(2021)Miguet, Rouyer, and Rio]{miguet2021life}
J.~Miguet, F.~Rouyer and E.~Rio, \emph{Molecules}, 2021, \textbf{26},
  1317\relax
\mciteBstWouldAddEndPuncttrue
\mciteSetBstMidEndSepPunct{\mcitedefaultmidpunct}
{\mcitedefaultendpunct}{\mcitedefaultseppunct}\relax
\EndOfBibitem
\bibitem[Lhuissier and Villermaux(2012)]{lhuissier2012bursting}
H.~Lhuissier and E.~Villermaux, \emph{Journal of Fluid Mechanics}, 2012,
  \textbf{696}, 5--44\relax
\mciteBstWouldAddEndPuncttrue
\mciteSetBstMidEndSepPunct{\mcitedefaultmidpunct}
{\mcitedefaultendpunct}{\mcitedefaultseppunct}\relax
\EndOfBibitem
\bibitem[Aradian \emph{et~al.}(2001)Aradian, Rapha{\"{e}}l, and
  de~Gennes]{Aradian2001}
A.~Aradian, E.~Rapha{\"{e}}l and P.-G. de~Gennes, \emph{Europhys. Lett.}, 2001,
  \textbf{55}, 834--840\relax
\mciteBstWouldAddEndPuncttrue
\mciteSetBstMidEndSepPunct{\mcitedefaultmidpunct}
{\mcitedefaultendpunct}{\mcitedefaultseppunct}\relax
\EndOfBibitem
\bibitem[Poulain \emph{et~al.}(2018)Poulain, Villermaux, and
  Bourouiba]{poulain2018ageing}
S.~Poulain, E.~Villermaux and L.~Bourouiba, \emph{Journal of fluid mechanics},
  2018, \textbf{851}, 636--671\relax
\mciteBstWouldAddEndPuncttrue
\mciteSetBstMidEndSepPunct{\mcitedefaultmidpunct}
{\mcitedefaultendpunct}{\mcitedefaultseppunct}\relax
\EndOfBibitem
\bibitem[Champougny \emph{et~al.}(2016)Champougny, Roch{\'{e}}, Drenckhan, and
  Rio]{Champougny2016}
L.~Champougny, M.~Roch{\'{e}}, W.~Drenckhan and E.~Rio, \emph{Soft matter},
  2016, \textbf{12}, 5276--5284\relax
\mciteBstWouldAddEndPuncttrue
\mciteSetBstMidEndSepPunct{\mcitedefaultmidpunct}
{\mcitedefaultendpunct}{\mcitedefaultseppunct}\relax
\EndOfBibitem
\bibitem[Frazier \emph{et~al.}(2020)Frazier, Jiang, and
  Burton]{frazier2020make}
S.~Frazier, X.~Jiang and J.~C. Burton, \emph{Physical Review Fluids}, 2020,
  \textbf{5}, 013304\relax
\mciteBstWouldAddEndPuncttrue
\mciteSetBstMidEndSepPunct{\mcitedefaultmidpunct}
{\mcitedefaultendpunct}{\mcitedefaultseppunct}\relax
\EndOfBibitem
\bibitem[Stubenrauch and Khristov(2005)]{stubenrauch2005foams}
C.~Stubenrauch and K.~Khristov, \emph{Journal of colloid and interface
  science}, 2005, \textbf{286}, 710--718\relax
\mciteBstWouldAddEndPuncttrue
\mciteSetBstMidEndSepPunct{\mcitedefaultmidpunct}
{\mcitedefaultendpunct}{\mcitedefaultseppunct}\relax
\EndOfBibitem
\bibitem[Cheng(2008)]{cheng2008formula}
N.-S. Cheng, \emph{Industrial \& engineering chemistry research}, 2008,
  \textbf{47}, 3285--3288\relax
\mciteBstWouldAddEndPuncttrue
\mciteSetBstMidEndSepPunct{\mcitedefaultmidpunct}
{\mcitedefaultendpunct}{\mcitedefaultseppunct}\relax
\EndOfBibitem
\bibitem[Volk and K{\"a}hler(2018)]{volk2018density}
A.~Volk and C.~J. K{\"a}hler, \emph{Experiments in Fluids}, 2018, \textbf{59},
  1--4\relax
\mciteBstWouldAddEndPuncttrue
\mciteSetBstMidEndSepPunct{\mcitedefaultmidpunct}
{\mcitedefaultendpunct}{\mcitedefaultseppunct}\relax
\EndOfBibitem
\bibitem[Boulogne(2019)]{Boulogne2019}
F.~Boulogne, \emph{The European Physical Journal E}, 2019, \textbf{42},
  51\relax
\mciteBstWouldAddEndPuncttrue
\mciteSetBstMidEndSepPunct{\mcitedefaultmidpunct}
{\mcitedefaultendpunct}{\mcitedefaultseppunct}\relax
\EndOfBibitem
\bibitem[Boulogne(2019--)]{Oospectro}
F.~Boulogne, \emph{{Oospectro}: Get thicknesses from Ocean Optics spectra},
  2019--, \url{https://github.com/sciunto-org/oospectro}\relax
\mciteBstWouldAddEndPuncttrue
\mciteSetBstMidEndSepPunct{\mcitedefaultmidpunct}
{\mcitedefaultendpunct}{\mcitedefaultseppunct}\relax
\EndOfBibitem
\bibitem[Miguet \emph{et~al.}(2021)Miguet, Pasquet, Rouyer, Fang, and
  Rio]{MiguetPRF2021}
J.~Miguet, M.~Pasquet, F.~Rouyer, Y.~Fang and E.~Rio, \emph{Phys. Rev. Fluids},
  2021, \textbf{6}, L101601\relax
\mciteBstWouldAddEndPuncttrue
\mciteSetBstMidEndSepPunct{\mcitedefaultmidpunct}
{\mcitedefaultendpunct}{\mcitedefaultseppunct}\relax
\EndOfBibitem
\bibitem[{J. Mysels S. Frankel}(1959)]{Frankel1959}
K.~S. {J. Mysels S. Frankel}, \emph{{Soap FIlms: studies of their thinning}},
  Pergamon Press, New York - London - Paris - Los Angeles, 1959, p. 114\relax
\mciteBstWouldAddEndPuncttrue
\mciteSetBstMidEndSepPunct{\mcitedefaultmidpunct}
{\mcitedefaultendpunct}{\mcitedefaultseppunct}\relax
\EndOfBibitem
\bibitem[Champougny \emph{et~al.}(2015)Champougny, Scheid, Restagno, Vermant,
  and Rio]{champougny2015surfactant}
L.~Champougny, B.~Scheid, F.~Restagno, J.~Vermant and E.~Rio, \emph{Soft
  matter}, 2015, \textbf{11}, 2758--2770\relax
\mciteBstWouldAddEndPuncttrue
\mciteSetBstMidEndSepPunct{\mcitedefaultmidpunct}
{\mcitedefaultendpunct}{\mcitedefaultseppunct}\relax
\EndOfBibitem
\bibitem[Seiwert \emph{et~al.}(2014)Seiwert, Dollet, and
  Cantat]{seiwert2014theoretical}
J.~Seiwert, B.~Dollet and I.~Cantat, \emph{Journal of fluid mechanics}, 2014,
  \textbf{739}, 124--142\relax
\mciteBstWouldAddEndPuncttrue
\mciteSetBstMidEndSepPunct{\mcitedefaultmidpunct}
{\mcitedefaultendpunct}{\mcitedefaultseppunct}\relax
\EndOfBibitem
\bibitem[Langevin(2014)]{langevin2014rheology}
D.~Langevin, \emph{Annual review of fluid mechanics}, 2014, \textbf{46},
  47--65\relax
\mciteBstWouldAddEndPuncttrue
\mciteSetBstMidEndSepPunct{\mcitedefaultmidpunct}
{\mcitedefaultendpunct}{\mcitedefaultseppunct}\relax
\EndOfBibitem
\bibitem[Jones \emph{et~al.}(2001--)Jones, Oliphant,
  Peterson,\emph{et~al.}]{Jones2001}
E.~Jones, T.~Oliphant, P.~Peterson \emph{et~al.}, \emph{{SciPy}: Open source
  scientific tools for {Python}}, 2001--, \url{http://www.scipy.org/}\relax
\mciteBstWouldAddEndPuncttrue
\mciteSetBstMidEndSepPunct{\mcitedefaultmidpunct}
{\mcitedefaultendpunct}{\mcitedefaultseppunct}\relax
\EndOfBibitem
\bibitem[Fuchs(1959)]{Fuchs1959}
N.~A. Fuchs, \emph{Evaporation and droplet growth in gaseous media}, Pergamon
  Press, 1959\relax
\mciteBstWouldAddEndPuncttrue
\mciteSetBstMidEndSepPunct{\mcitedefaultmidpunct}
{\mcitedefaultendpunct}{\mcitedefaultseppunct}\relax
\EndOfBibitem
\bibitem[Bouchaudy \emph{et~al.}(2018)Bouchaudy, Loussert, and
  Salmon]{Bouchaudy2018}
A.~Bouchaudy, C.~Loussert and J.-B. Salmon, \emph{AIChE Journal}, 2018,
  \textbf{64}, 358--366\relax
\mciteBstWouldAddEndPuncttrue
\mciteSetBstMidEndSepPunct{\mcitedefaultmidpunct}
{\mcitedefaultendpunct}{\mcitedefaultseppunct}\relax
\EndOfBibitem
\bibitem[Association
  \emph{et~al.}(1963)Association\emph{et~al.}]{glycerine1963physical}
G.~P. Association \emph{et~al.}, \emph{Physical properties of glycerine and its
  solutions}, Glycerine Producers' Association, 1963\relax
\mciteBstWouldAddEndPuncttrue
\mciteSetBstMidEndSepPunct{\mcitedefaultmidpunct}
{\mcitedefaultendpunct}{\mcitedefaultseppunct}\relax
\EndOfBibitem
\bibitem[Chen \emph{et~al.}(2009)Chen, Li, Song, and Yang]{chen2009hydrogen}
C.~Chen, W.~Z. Li, Y.~C. Song and J.~Yang, \emph{Journal of Molecular Liquids},
  2009, \textbf{146}, 23--28\relax
\mciteBstWouldAddEndPuncttrue
\mciteSetBstMidEndSepPunct{\mcitedefaultmidpunct}
{\mcitedefaultendpunct}{\mcitedefaultseppunct}\relax
\EndOfBibitem
\bibitem[Roux \emph{et~al.}(2022)Roux, Duchesne, and Baudoin]{roux2021bubbles}
A.~Roux, A.~Duchesne and M.~Baudoin, \emph{Phys. Rev. Fluids}, 2022,
  \textbf{7}, L011601\relax
\mciteBstWouldAddEndPuncttrue
\mciteSetBstMidEndSepPunct{\mcitedefaultmidpunct}
{\mcitedefaultendpunct}{\mcitedefaultseppunct}\relax
\EndOfBibitem
\bibitem[La~Mer(2014)]{la2014retardation}
V.~K. La~Mer, \emph{Retardation of evaporation by monolayers: transport
  processes}, Academic Press, 2014\relax
\mciteBstWouldAddEndPuncttrue
\mciteSetBstMidEndSepPunct{\mcitedefaultmidpunct}
{\mcitedefaultendpunct}{\mcitedefaultseppunct}\relax
\EndOfBibitem
\bibitem[Li \emph{et~al.}(2010)Li, Shaw, and Stevenson]{li2010effect}
X.~Li, R.~Shaw and P.~Stevenson, \emph{International Journal of Mineral
  Processing}, 2010, \textbf{94}, 14--19\relax
\mciteBstWouldAddEndPuncttrue
\mciteSetBstMidEndSepPunct{\mcitedefaultmidpunct}
{\mcitedefaultendpunct}{\mcitedefaultseppunct}\relax
\EndOfBibitem
\end{mcitethebibliography}
\bibliographystyle{rsc} 

\end{document}